\documentclass[apj]{emulateapj}
\usepackage{graphics,graphicx}
\usepackage{epsfig}
\usepackage{dcolumn}
\usepackage{rotating}
\usepackage{natbib}
\shorttitle{A panchromatic view of PKS~0558--504}
\shortauthors{Gliozzi et al.}

  \def\pks{PKS~0558--504}

  \def\xmm{{\it XMM-Newton}} 
   
  \def\atca{{\it ATCA}}
  
  \def\swift{{\it SWIFT}} 
  \def\rxte{{\it RXTE}}

  \def\lum{erg s$^{-1}$}

  \def\arcsec{$^{\prime\prime}$}

  \def\ltsima{$\; \buildrel < \over \sim \;$}
  \def\simlt{\lower.5ex\hbox{\ltsima}} 
  \def\gtsima{$\; \buildrel > \over \sim \;$}
  \def\simgt{\lower.5ex\hbox{\gtsima}} 

\begin{document}
\title{A panchromatic view of PKS~0558--504: an ideal laboratory to study the disk-jet link}

\author{M. Gliozzi}
\affil{George Mason University, 4400 University Drive, Fairfax, VA 22030}

\author{I.E. Papadakis}
\affil{Physics Department, University of Crete, 710 03 Heraklion,
Crete, Greece}
\affil{Foundation for Research and Technology - Hellas,
IESL, Voutes, 71110 Heraklion, Crete, Greece}

\author{D. Grupe}
\affil{Department of Astronomy and Astrophysics, Pennsylvania State University, 525 Davey Lab, University Park, PA 16802, USA}

\author{W.P. Brinkmann, C. Raeth}
\affil{Max-Planck-Institut f\"ur 
extraterrestrische Physik, Postfach 1312, D-85741 Garching, Germany}

\author{L. Kedziora-Chudczer} 
\affil{Australia Telescope National Facility, CSIRO, P.O. Box 76, Epping, 
NSW 1710 {\it Currently at the School of Physics, UNSW, Sydney, NSW 2052 Australia}}

\begin{abstract}
PKS 0558­-504 is the brightest radio-loud Narrow-Line Seyfert 1 galaxy 
at X-ray energies. Here we present results from the radio, optical, UV,
and X-ray bands obtained with \swift, \xmm, and \atca\ during a 
10-day monitoring campaign in September 2008. The simultaneous coverage 
at several wavelengths makes it possible to investigate in detail
the broadband spectral energy distribution (SED) and
the energetic of this source. The main results can be summarized as follows.
The \atca\ reveals the presence of an extended radio emission in PKS~0558--504 
with two lobe-like structures at $\sim$7\arcsec\ from the bright central source.
The extended radio structure and the low value of the radio-loudness
similar to radio-quiet Seyfert galaxies coupled with constraints from
higher energy bands argue against a jet-dominated emission in \pks.
The study of the SED, which is dominated by a nearly constant optical-UV emission, 
supports the conclusion that PKS~0558--504 is accreting at super-Eddington rate. 
This conclusion was reached assuming $M_{\rm BH}=2.5\times10^8~{M_\odot}$, which
was obtained with a new scaling method based on X-ray spectral variability results. 
A comparison between the accretion luminosity and the kinetic power associated 
with the jet suggests that in this source the accretion power dominates  
in agreement with the results obtained from Radiation-MHD 
simulations of Galactic black holes (GBHs) accreting at the Eddington rate. The 
combined findings from this panchromatic investigation strongly suggest that \pks\ is a
large scale analog of GBHs in their highly accreting intermediate state. Importantly,
\pks\ may also be the prototype of the parent population of the very radio-loud
NLS1s recently detected at $\gamma$-ray energies.
\end{abstract}


\section{Introduction}
Bipolar relativistic jets are common 
features in a variety of astrophysical objects, most notably in 
Galactic Black Holes (GBHs) and Active Galactic Nuclei (AGNs). 
Accretion of gas onto black holes is 
thought to power these collimated outflows. However, the details of the
jet formation as well as the nature of the coupling between accreting
matter and outflows are still among the outstanding open questions
in high energy astrophysics.

Because of their vicinity and hence their high brightness, the temporal and spectral 
properties of GBHs are much better known and can be used to
infer information on their more powerful, extragalactic analogs.
Indeed considerable progress in this field has been made by multi-wavelength 
correlated studies of GBHs in different spectral states, which allow one to 
study the link between accretion (generally probed by the X-ray emission) 
and jet properties (in the radio regime) on ``human'' timescales. 
It is now well established that GBHs undergo state
transitions, switching between two main states: the low/hard (LS) and 
the high/soft state (HS) passing through  soft and hard intermediate states 
(IS), which are also named very high states (VHS) when they occur at high 
values of accretion rate 
(see \citealt{mcclin06,done07} for recent comprehensive reviews on GBHs). 
Each spectral state is
unambiguously characterized 
by a specific combination of X-ray temporal and spectral properties and by 
well defined radio features \citep[see][]{fend04}.

In the study of the disk-jet link
one of the most interesting spectral states is the IS/VHS, which 
is generally characterized by powerful transient relativistic ejections in the radio
coupled with highly variable X-ray emission that is unambiguously associated 
with the accretion flow, unlike the LS where the origin of the X-rays is 
still matter of debate \citep[e.g.][]{mark03,zdzi04}.

Unfortunately, the physical conditions that lead to the transient 
relativistic ejections during the IS/VHS
are still poorly understood mostly because of their short duration in GBHs. 
Since the 
dynamical time scales are proportional to the black hole mass, in individual 
AGNs it is not possible to observe 
long-term phenomena occurring in GBHs such as the canonical spectral 
transitions. On the other hand, AGNs may provide better constraints on 
short-lived GBH phenomena and hence shed light on
the jet formation and the interplay between accretion and ejection
processes. 
\begin{table*}[bht]
\scriptsize
\caption{\swift\ Observation log of PKS 0558-504}
\begin{center}
\begin{tabular}{lcccrrrrrrr}
\hline        
\hline
\noalign{\smallskip} 
Segment & Start time  & End Time & MJD & \multicolumn{7}{c}{Observing time given in s}\\
 &(UT) &(UT) & & $\rm T_{\rm XRT}$& $\rm T_{\rm V}$ & $\rm T_{\rm B}$ &$\rm T_{\rm U}$ & $\rm T_{\rm UVW1}$ &
$\rm T_{\rm UVM2}$ & $\rm T_{\rm UVW2}$\\
\noalign{\smallskip}
\hline
\noalign{\smallskip}
001 & 2008-09-07 08:10 & 2008-09-07 11:36 & 54716.41 & 2215 & 122 & 122 & 122 & 244 & 210 & 488 \\
002 & 2008-09-08 01:50 & 2008-09-08 06:50 & 54717.20 & 2429 & 199 & 199 & 199 & 399 & 358 & 798 \\
003 & 2008-09-09 09:52 & 2008-09-09 11:48 & 54718.45 & 2327 & 191 & 191 & 191 & 381 & 532 & 763 \\
004 & 2008-09-10 00:15 & 2008-09-10 18:20 & 54719.39 & 2203 & 181 & 181 & 181 & 361 & 509 & 724 \\
005 & 2008-09-11 13:21 & 2008-09-11 19:59 & 54720.69 & 1897 & 163 & 163 & 163 & 326 & 344 & 653 \\
006 & 2008-09-12 00:36 & 2008-09-12 05:37 & 54721.15 & 2008 & 155 & 155 & 155 & 310 & 483 & 621 \\
007 & 2008-09-13 15:09 & 2008-09-13 18:40 & 54722.68 & 1869 & 148 & 148 & 148 & 296 & 447 & 594 \\
008 & 2008-09-14 00:42 & 2008-09-14 05:40 & 54723.13 & 2363 & 240 & 240 & 240 & 483 & 682 & 966 \\
009 & 2008-09-15 16:56 & 2008-09-15 21:54 & 54724.80 & 2353 & 184 & 184 & 184 & 369 & 557 & 739 \\
010 & 2008-09-16 07:20 & 2008-09-16 12:11 & 54725.90 & 2072 & 165 & 165 & 165 & 330 & 461 & 660 \\
\noalign{\smallskip}
\hline        
\hline
\end{tabular}
\end{center}
\label{tab1}
\end{table*} 

In the framework of the AGN-GBH unification, the Narrow-Line 
Seyfert 1 galaxies (NLS1s) are the best candidates for large-scale analogs of 
GBHs in the IS/VHS.
NLS1s are historically identified by their optical 
emission 
line properties: the ratio [O III]/H$\beta$ is less than 3 and FWHM H$\beta$ is
less than 2000${~\rm km~s^{-1}}$ \citep{oster85,good89}.
They are seldom radio loud \citep{komo06}, although recent studies reveal 
the existence of several NLS1s characterized by very high radio-loudness 
\citep[e.g.][]{yuan08}. Recently, a few of these very radio-loud NLS1s have 
been detected at $\gamma$-ray
energies by the {\it Fermi/LAT} collaboration confirming that these sources posses
relativistic jets observed at small viewing angles \citep{abdo09a,abdo09b,abdo09c,
fosc09}. In the X-rays, NLS1 are generally characterized by steep spectra and 
strong variability \citep[e.g.][]{brand99,lei99a,lei99b,gru01}. Based on these 
properties, it has been suggested that NLS1 are AGN in their early phase 
\citep{gru99,math00}, characterized by relatively small black 
hole masses \citep[e.g.][]{gru04}, and very high accretion rates in 
terms of Eddington units 
(e.g. \citealt{boro92,sule00,gru10}; see also \citealt{marco08,deca08}
for a discording view).

PKS~0558--504 ($z=0.137$) is the brightest radio-loud 
($R=L_\nu(6~{\rm cm})/L_\nu({\rm B})\simeq 27$, \citealt{sieb99})
NLS1 in the X-ray band
and therefore one of the best studied at these energies.
Previous X-ray observations with different satellites have confirmed 
that \pks\ shows the characteristic NLS1 properties: strong variability,
steep X-ray spectrum, substantial soft excess, and relatively high luminosity. 
By comparing the X-ray observations from several satellites over 
a decade, it is evident that on long timescales
the strong X-ray variability of PKS~0558--504 occurs 
persistently but without being accompanied by significant spectral
variability: the spectrum above 2 keV is consistently described by a
power-law model with photon index  $\Gamma\sim2.2$ \citep{glioz00}.
This conclusion  has been confirmed by a recent monitoring campaign 
with \rxte: the long-term achromatic variability appears to be consistent with the 
behavior of GBHs during the transition LS-to-HS and inconsistent
with the spectral variability typical of jet-dominated AGNs \citep{glioz07}. 

In the last ten years,
PKS~0558--504 has been observed repeatedly by \xmm\ \citep{obri01,glioz01,brink04}.
The most recent and highest
quality data, obtained with the \xmm\ EPIC cameras from five consecutive 
orbits in September 2008, confirm that \pks\ is highly variable on all 
sampled timescales, and a spectral analysis indicates that the 2--10 keV 
energy range is well fitted with
a power law whose slope steepens as the source brightens and the soft excess
is well described by a low-temperature Comptonization model \citep{papa10a}. 
These findings suggest that despite its radio loudness \pks\ appears to
behave like a ``normal'' Seyfert galaxy from the X-ray point of view.

The main goal of this work is to shed light on the energetic of this 
powerful source using multi-wavelength data. For this purpose,
after describing the observations and data reduction in $\S~2$, we
first try to assess the role played by the jet emission 
beyond the radio band (in $\S~3$) and then to 
constrain the black hole mass in \pks\ without the use of optical measurements
(in $\S~4$). In Section 5 we describe the 
multi-wavelength behavior of \pks\ based on a 10-day multi-wavelength campaign 
carried out with the \swift\ XRT and UVOT simultaneously with the deep \xmm\ 
observations in September 2008, and complemented with one {\it radio} observation with 
the Australia Telescope Compact Array (ATCA). Finally,
in $\S~6$ we discuss the implications of the main results and draw our conclusions.
Hereafter, we adopt a cosmology with $H_0=71{\rm~km~s^{-1}~Mpc^{-1}}$,
$\Omega_\Lambda=0.73$ and $\Omega_{\rm M}=0.27$ \citep{ben03}; with 
the assumed cosmological parameters, the luminosity
distance of \pks\  is 642 Mpc.

\section{Observations and Data Reduction}

\subsection{SWIFT}
PKS 0558-504 was observed by the \swift\ Gamma-Ray Burst  explorer
mission  \citep{ger04} between 2008 September 9 - 16. The details of this
intense monitoring campaign are summarized in Tables 1 and 2. The \swift\
X-ray Telescope (XRT; \citealt{burro05}) observations were all performed in
Windowed Timing mode (WT; \citealt{hill04}) in order to avoid the effects of
pile-up. Data were reduced by the task {\it xrtpipeline} version 0.12.1., 
which is included in the HEASOFT package 6.8. 
Source and background photons were selected in boxes 40 pixel long.
Only single and double events (GRADES 0 to 2) were used. 
Source photons for the light curve and spectra were extracted  
with {\it XSELECT}.  
The auxiliary response files
(ARFs) were created using {\it xrtmkarf} version 0.5.6 and the response matrix 
{\it swxwt0to2s6\_20010101v011.rmf}.

The UV/Optical Telescope (UVOT; Roming et al. 2005) observed PKS 0558-504 in all 6 filters.
The exposure times in each of the filters per segment are given in Table 1 and the
corresponding magnitudes in Table 2.
The UVOT data  were reduced and analyzed as
described in  \citet{pool08}.  
Source photons were extracted from the coadded images files for each segment
with a radius of
$5\farcs$ Backgound photons were selected in a nearby source-free region with
r=$20\farcs$ Magnitudes and fluxes were measured using the UVOT tool {\it
uvotsource}. All magnitudes listed in Table 2 were corrected for Galactic reddening
($E_{\rm B-V}$=0.044; \citealt{sfd98}). The correction factors were based on the standard
reddening correction curves by \citet{card89}  as described by equation 2 in
\citet{rom09}.

\subsection{XMM-Newton}
\pks\ was observed by \xmm\ from September 9, 2008, 1:19 UT to September 16,
2008, 12:02 UT. The EPIC pn and MOS1 cameras were operated in small window 
mode, the MOS2 in timing mode, and both RGS in spectroscopy mode. In the 
present work we only use EPIC pn data, which were processed with 
the \xmm\ Science Analysis Software (\verb+SAS+) 8.0. 
The recorded single and double events were screened to remove known
hot pixels and other data flagged as bad: only data with {\tt PATTERN$\leq$4,
FLAG=0} were used. A more detailed description of the \xmm\ data analysis
can be found in \citet{papa10a}.

\subsection{ATCA}
\begin{figure}
\includegraphics[bb=24 190 565 625,clip=,angle=0,width=8.5cm]{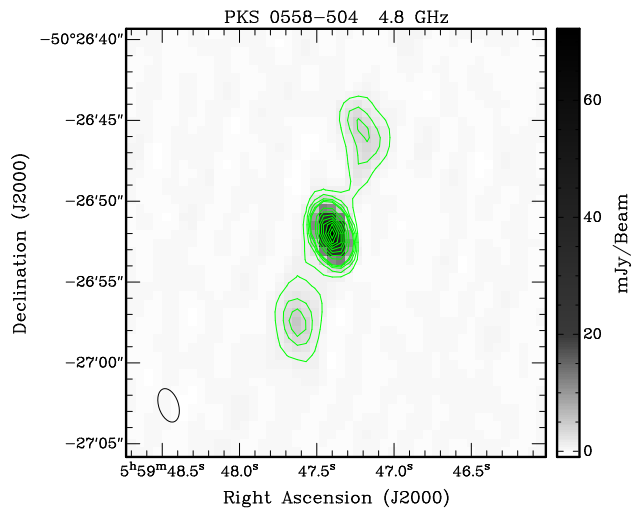}
\caption{\scriptsize
The {\it ATCA} image with the rms of $2\times10^{-4}$ Jy/beam was obtained on the 19th of January 2008. The superimposed contour lines show isophotes between 2\% and 95\% of the peak flux density.  
The beam size is shown in the bottom left corner. }
\label{figure:fig1}
\end{figure}

Radio monitoring observations of the \pks\ with the ATCA commenced in April 2006. The observation session
that overlapped with the \swift\ and \xmm\ monitoring campaign was carried out on 
September 16, 2008, 22:50 UT.
The source showed a slightly inverted spectrum between 18.5 and 4.8 GHz
($\alpha\simeq0.3$, where the flux density $f_\nu$ is related to the spectral 
index $\alpha$ by the relation
$f_\nu\propto \nu^{-\alpha}$). The peak flux 
density of 0.09$\pm$0.01 Jy  was measured at 4.8 GHz.
Our previous images at this frequency indicate that almost 80\% of radio intensity 
originates from the bright unresolved core, as shown in Figure~\ref{figure:fig1}.  
No flux density fluctuations were detected on intra-hourly timescales. 

\subsection{SED fitting}
The broadband spectral analysis  was performed using the {\tt XSPEC v.12.4}
software package \citep{arn96}.  We used {\tt FLX2XSP} in {\tt FTOOLS} to 
transform the optical and UV fluxes into suitable units for XSPEC.
All the X-ray spectra were re-binned so 
that each bin contained at least 20 counts for the
$\chi^2$ statistic to be valid. 

\begin{table*}
\scriptsize
\caption{\swift~XRT count rates and hardness ratios and UVOT Magnitudes of PKS 0558-504}
\begin{center}
\begin{tabular}{lcccccccc}
\hline        
\hline
\noalign{\smallskip}
Segment & XRT rate & XRT HR & V & B & U & UVW1 & UVM2 & UVW2 \\
\noalign{\smallskip}
\hline
\noalign{\smallskip}
001 & 0.747$\pm$0.020 & --0.11$\pm$0.02  & 14.90$\pm$0.03 & 15.06$\pm$0.02 & 13.83$\pm$0.01 & 13.54$\pm$0.01 & 13.23$\pm$0.02 & 13.29$\pm$0.01 \\
002 & 1.406$\pm$0.025 & --0.05$\pm$0.02  & 14.85$\pm$0.02 & 15.07$\pm$0.01 & 13.85$\pm$0.01 & 13.55$\pm$0.01 & 13.22$\pm$0.01 & 13.30$\pm$0.01 \\
003 & 0.933$\pm$0.021 & --0.12$\pm$0.02  & 14.98$\pm$0.02 & 15.08$\pm$0.01 & 13.85$\pm$0.01 & 13.63$\pm$0.01 & 13.36$\pm$0.01 & 13.43$\pm$0.01 \\
004 & 0.856$\pm$0.025 & --0.05$\pm$0.03  & 14.90$\pm$0.03 & 15.07$\pm$0.02 & 13.89$\pm$0.01 & 13.62$\pm$0.01 & 13.33$\pm$0.01 & 13.33$\pm$0.01 \\
005 & 0.808$\pm$0.022 & --0.15$\pm$0.03  & 14.84$\pm$0.02 & 15.05$\pm$0.01 & 13.84$\pm$0.01 & 13.57$\pm$0.01 & 13.23$\pm$0.01 & 13.33$\pm$0.01 \\
006 & 1.080$\pm$0.025 & --0.11$\pm$0.02  & 14.84$\pm$0.02 & 15.05$\pm$0.01 & 13.86$\pm$0.01 & 13.53$\pm$0.01 & 13.25$\pm$0.01 & 13.30$\pm$0.02 \\
007 & 1.010$\pm$0.025 & --0.07$\pm$0.02  & 14.88$\pm$0.02 & 15.02$\pm$0.01 & 13.85$\pm$0.01 & 13.54$\pm$0.01 & 13.24$\pm$0.01 & 13.30$\pm$0.01 \\
008 & 0.777$\pm$0.019 & --0.07$\pm$0.02  & 14.89$\pm$0.02 & 15.02$\pm$0.01 & 13.84$\pm$0.01 & 13.55$\pm$0.01 & 13.23$\pm$0.01 & 13.31$\pm$0.01 \\
009 & 0.577$\pm$0.017 & --0.13$\pm$0.03  & 14.86$\pm$0.02 & 15.05$\pm$0.01 & 13.83$\pm$0.01 & 13.56$\pm$0.01 & 13.20$\pm$0.01 & 13.30$\pm$0.01 \\
010 & 0.840$\pm$0.022 & --0.16$\pm$0.03  & 14.87$\pm$0.02 & 15.04$\pm$0.01 & 13.87$\pm$0.01 & 13.58$\pm$0.01 & 13.23$\pm$0.01 & 13.31$\pm$0.01 \\
\noalign{\smallskip}
\hline        
\hline
\end{tabular}
\end{center}
\label{tab2}
\scriptsize{
The magnitude was corrected for reddening with $E_{\rm B-V}$=0.044 given by
\citet{sfd98}. The errors given in this table are statistical errors.}
\end{table*}

\section{Radio Properties of PKS~0558-504}
Despite the fact that \pks\ has been one of the first radio-loud NLS1 ever 
discovered, not until this project its radio properties and activity have been studied in deserving detail. 
Our multi-wavelength, radio monitoring and imaging program with the {\it ATCA} and the VLBI 
shows the long-term variability pattern similar to the flat spectrum, low luminosity compact radio sources \citep{falc00}. 
A detailed description of the radio properties will be presented elsewhere (Kedziora-Chudczer et al. 2010, in 
preparation). Here we limit ourselves to briefly discuss the morphology of the radio emission and the intensity of the source during the \swift\ and \xmm\ campaigns in September 2008. 

From Figure~\ref{figure:fig1} it is evident that the radio emission of \pks\ resembles an FRI radio galaxy that is 
characterized by two diffuse lobes located at $\sim$ 7\arcsec\ from the bright, central core. At the redshift of $z=0.137$ the 
projected linear size of the full structure is close to 46 kpc. Interestingly, the
extended radio emission observed in \pks\ is at odds with the compact radio structure 
generally observed in other radio-loud NLS1s, which are characterized by 
radio-loudness parameters larger than 100 \citep[e.g.][]{yuan08}.

Historically, the radio-loudness parameter has been defined as 
$R\equiv L_\nu(6~{\rm cm})/L_\nu({\rm B})$  with $R=10$ considered as the conventional boundary between radio-loud and radio-quiet AGNs \citep[e.g.][]{kell94}. However,
\citet{hopeng01} demonstrated that, when the nuclear luminosities are properly measured, most
of the Seyfert galaxies appear to be radio-loud, indicating that $R=10$ is not a reliable
boundary. This conclusion is further supported by the recent findings of \citet{siko07},
who showed that radio-loud and radio-quiet AGNs form two distinct
branches when the radio-loudness is plotted versus the Eddington ratio $\lambda_{\rm Edd}$.
Importantly, while the two branches are nearly horizontal and well separated at
low values of the Eddington ratio, as $\lambda_{\rm Edd}$ increases
the branch slopes become negative and the branches broaden with substantial overlap 
at large values of the Eddington ratio.  
Unfortunately, the super-Eddington accretion rate in \pks\ 
(see Sect. 5.2)
coupled with the lack of a clear distinction between the two branches
at large values of the Eddington ratio hampers the study of the radio-loudness for \pks\
using the $\log R - \log \lambda_{\rm Edd}$ plot.

\begin{figure}
\includegraphics[bb=55 30 355 300,clip=,angle=0,width=8.5cm]{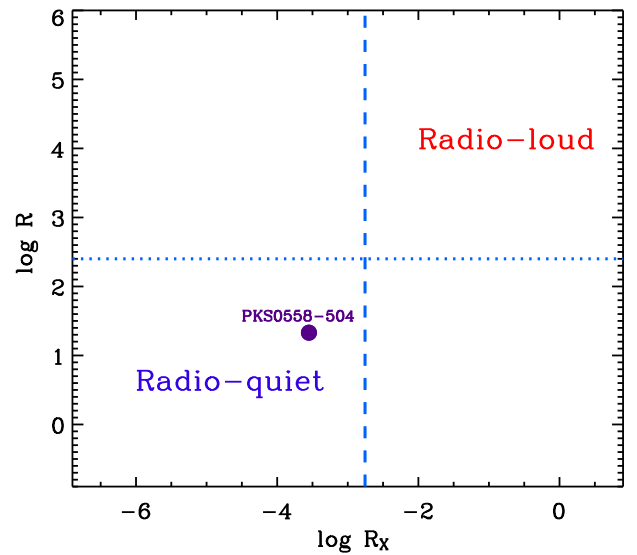}
\caption{\scriptsize
Classical radio loudness 
$R$ plotted versus the X-ray radio loudness $R_{\rm x}$. \pks\ is well
within the radio-quiet region and apparently inconsistent with the radio-loud
objects.The 
boundaries between radio-loud and radio-quiet objects have been determined
by \citet{panes07} using two large samples of radio-quiet Seyfert galaxies
and LLRGs.}
\label{figure:fig2}
\end{figure}

Nevertheless, the radio-loudness of \pks\ can be quantitatively assessed by comparing it
with the findings of \citet{panes07} who carried out a detailed investigation of  the 
radio-loudness in two large samples of radio-quiet 
Seyfert galaxies and Low-Luminosity Radio Galaxies (LLRGs). In this study both the 
classical radio loudness parameter $R$ and the X-ray radio loudness 
$R_{\rm x}\equiv\nu L_\nu(6~{\rm cm})/L_{\rm 2-10~ keV}$ were used; the latter 
parameter was introduced by \citet{tera03}
to circumvent extinction problems that usually affect the optical
emission and may lead to overestimates of $R$. Using nearly simultaneous radio, optical,
and X-ray observations (we have used the flux values measured 
during the last  day of the \xmm\ - \swift\ campaign because they are almost contemporaneous
with the \atca\ observation) for the $\log R - \log R_{\rm x}$ plot,  \pks\ appears to be 
fully consistent with radio-quiet Seyfert galaxies and
inconsistent with radio-loud objects, as visually demonstrated in
Fig~\ref{figure:fig2}, where the boundaries have been determined by 
\citet{panes07}.

In conclusion, the extended radio emission and the symmetric location of the 
lobes in \pks, which are reminiscent of the typical structure observed in FRI radio 
galaxies, argue against a highly beamed source. This is confirmed by the simultaneous 
radio-loudness parameters that are in full agreement with radio-quiet Seyfert galaxies.
On the other hand, our VLBI imaging at 2.3 GHz 
(Kedziora-Chudczer et al. 2010, in preparation), which shows a one-sided pc-scale jet,
suggests that the presence of beamed emission cannot be ruled out for \pks\
in the radio band.  For the sake of clarity and for historical reasons in the reminder 
of the paper we will maintain the classification of radio-loud NLS1 for \pks.

\section{Black Hole Mass of PKS 0558-504}
In order to determine the accretion rate of \pks\ and investigate its
energetics, it is first necessary to constrain the mass of the supermassive
black hole. Here, we report four different and independent  measurements of 
$M_{\rm BH}$ for \pks.

{$\bullet$} (1) As for many AGNs for which there are no direct measurements from reverberation
mapping, the $M_{\rm BH}$ in  \pks\ has been estimated from the virial relationship
$M_{\rm BH}= f R(\Delta V)^2/G$ (where $f$ is an unknown geometric factor, and
$R$ and $(\Delta V)^2$ are the radius 
and velocity dispersion of broad line region, respectively). This yielded 
$M_{\rm BH}\simeq 6 \times 10^7 ~M_\odot$ \citep{papa10a}, which is relatively 
small when compared to the typical masses of radio-loud 
AGNs \citep[e.g.][]{mclur04}, but fully consistent 
with the values derived in NLS1s with the same method
\citep[e.g.][]{gru04}.

Given that the optical measurements --$L_{5100 \AA}$ \citep{corb00} and 
$H_\beta$ \citep{corb97}-- are non-simultaneous, and given the uncertainty on 
the geometric factor $f$ and the controversy about the application of the virial method to NLS1s (see e.g. \citealt{marco08,deca08}; but also \citealt{netz09}),
it is important to constrain $M_{\rm BH}$ also with alternative 
techniques that are independent of optical measurements and any assumption on the
BLR.

{$\bullet$} (2) One possible alternative method is based on the so called ``fundamental plane" of black
holes (BHs) introduced by \citet{merlo03} and  \citet{falc04}, 
where $M_{\rm BH}$ is related to both the X-ray and radio luminosities in any
accreting BH systems. Recently this relationship has been refined by \citet{guelt09}
 by utilizing only sources with BH masses that have been determined
dynamically. Using the latter relationship in combination of quasi simultaneous 
measurements of the radio and 
X-ray emission in \pks\
we derive $M_{\rm BH}\simeq 3\times 10^8 ~M_\odot$, which is 
larger than the value obtained from the virial theorem
by a factor of $\sim$4. This finding does not depend on the nearly simultaneous
nature of the observations, since we derive a very similar value for $M_{\rm BH}$ 
by using radio and \rxte\ luminosities averaged over an interval of one year. 

{$\bullet$} (3) For AGNs with evenly-sampled long-term coverage in the X-ray
band, a viable technique to estimate the mass of the black hole is based on the 
relationship $\log M_{\rm BH}= (\log T_{\rm B} + 0.98 \log L_{\rm bol} + 2.32)/2.1$,
where $M_{\rm BH}$ is in ${\rm 10^6~M_\odot}$ units,  $L_{\rm bol}$ is 
the bolometric luminosity in units of ${\rm 10^{44}~erg~s^{-1}}$, and $T_{\rm B}$
the time break in days derived from power spectral density (McHardy et al. 2006).
\pks\ has been regularly monitored with \rxte\ since March 2004, making it possible
to estimate $M_{\rm BH}$ with the above formula. The main
findings of a detailed temporal study obtained combining the long-term \rxte\ light curve
and the deep \xmm\ observation in September 2008 are reported by \citet{papa10b}
and suggest that $M_{\rm BH}\simeq 2-3\times 10^8 ~M_\odot$.

{$\bullet$} (4) An additional independent way to determine the mass in BH systems relies on the fact 
that hard X-rays are produced by the Comptonization process in both
stellar and supermassive black holes. Recently, \citet{shap09}
discovered that GBHs
present a universal scalable relationship between the photon index
and the normalization of the Bulk Motion Comptonization (BMC) model
during their spectral transitions. They also demonstrate that this relationship 
can be used to estimate the mass of any GBHs by simply scaling the $M_{\rm BH}$ 
value from a suitable reference source. We have started testing the 
extension of this method to SMBHs and the encouraging results will be published
elsewhere (Gliozzi et al. 2010 in preparation). Here we apply this method to \pks\
after describing the basic characteristics of the BMC model and of this scaling
technique.

 Although it was historically developed to 
describe the Comptonization of thermal seed photons by a relativistic converging 
flow \citep{tita97}, the BMC model is a generic Comptonization 
model able to describe equally well the thermal Comptonization (i.e., the 
inverse Compton scattering produced by electrons with a Maxwellian energy 
distribution) and the bulk motion Comptonization (where the seed photons are 
scattered off electrons with bulk relativistic motion). The BMC model is 
characterized by 4 free parameters: 1) the temperature of the
thermal seed photons $kT$, 2) the energy spectral index $\alpha$ (which is
related to the photon index by the relation $\Gamma=1+\alpha$), 3) a parameter 
$\log(A)$ related to the Comptonization fraction $f$ (.i.e., the ratio
between the number of Compton scattered photons and the number of seed 
photons) by the relation $f=A/(1+A)$, and 4) the normalization $N_{\rm BMC}$.

\begin{figure}
\includegraphics[bb=55 30 355 300,clip=,angle=0,width=8.5cm]{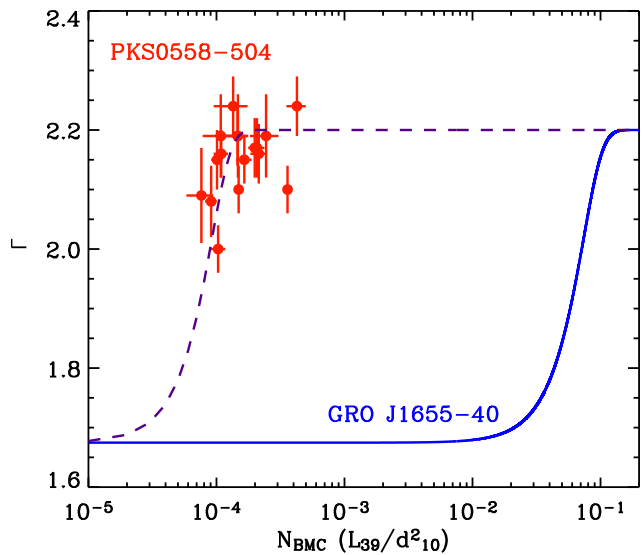}
\caption{\scriptsize
Photon index $\Gamma$ plotted versus the normalization of the 
Comptonization model N$_{\rm BMC}$. The scaling of the black hole mass
is provided by the horizontal shift between the 
\pks\ data points and the curve of the microquasar GRO~J1655-40}
\label{figure:fig3}
\end{figure}

In simple terms, the necessary steps to derive $M_{\rm BH}$ with  
this method can be summarized as follows:\\
\noindent 1) Construct a $\Gamma -
N_{\rm BMC}$ plot for a GBH of known mass and distance, which will be
used as reference (hereafter denoted by the subscript {\it r}).\\
\noindent 2) Compute the normalization ratio between the target of interest
(hereafter denoted by the subscript {\it t}) and the reference object
$N_{\rm BMC,t}/N_{\rm BMC,r}$ at the value of
$\Gamma$ measured for the AGN.\\
 \noindent 3) Derive the unknown black hole mass using the following equation
\begin{equation}
M_{\rm BH,t}=M_{\rm BH,r}\times 
(N_{\rm BMC,t}/N_{\rm BMC,r})\times (d_t/d_r)^2 \times f_G
\end{equation}
where $M_{\rm BH,r}$ is the black hole mass of the GBH reference object,
$N_{\rm BMC,t}$ and $N_{\rm BMC,r}$ are the respective BMC normalizations
for target and reference objects, $d_t$ and $d_r$ are the
corresponding distances, and $f_G=\cos\theta_r/\cos\theta_t$ is a
geometrical factor that depends on the respective inclination angles.

The above formula is readily obtained by considering that (a)
the normalization is a function of luminosity and distance: $N_{\rm
BMC}\propto L/d^2$, and (b) the luminosity of an accreting BH system
can be expressed by $L\propto
\eta M_{\rm BH} \dot m$, where $\eta$ is the radiative efficiency.
The only assumptions are that different sources in the same spectral state 
have similar values of $\eta$ and $\dot m$, and that
the photon index is a reliable indicator for
the spectral state. This is
in broad agreement with the positive correlation between photon index and
accretion rate found in bright AGNs \citep[e.g][]{shem06,papa09}.

Figure~\ref{figure:fig3} shows the $\Gamma - N_{\rm BMC}$ diagram for 
\pks\ in comparison with the primary reference source
GRO J1655-40, a well known microquasar whose parameters are 
tightly constrained: $M_{\rm BH}/M_{\odot}=6.3\pm0.3$, $i=70^o\pm1^o$,
$d=3.2\pm0.2$ kpc \citep{gree01,hjel95}. For
\pks\ we fitted the 2--10 keV energy band of the 7 relatively long 
segments (with exposures up to 20 ks)
that best describe the spectral evolution of the source during
the September 2008 \xmm\ campaign \citep{papa10a} and the 10 short
segments (2 ks exposures)
simultaneous to the \swift\ observation, which will be described 
below. For the spectral fits, the temperature of the thermal seed photons 
was kept frozen at the best fit value  obtained from the broadband
SED fit ($kT=8-23$ eV; see $\S5.2$), whereas the other parameters were free to vary.
The 2--10 keV spectra of
all segments were adequately fitted (typically $\chi_{\rm red} \simeq 0.9-0.95$)
with one BMC model absorbed by Galactic $N_{\rm H}$.

As expected, due to the longer timescales associated with SMBHs,
the \pks\ trend in the $\Gamma - N_{\rm BMC}$ diagram 
is restricted to a small portion of the trend shown by GRO J1655-40 during its 
2005 spectral evolution between LS and HS. Interestingly,
the short-term spectral behavior of \pks\ appears to be consistent with  
GRO J1655-40 in its highly accreting state. 

Substituting in  Eq. (1) a distance of 624 Mpc
and $N_{\rm BMC,t}/N_{\rm BMC,r}=8\times10^{-2}$  obtained from 
Fig.~\ref{figure:fig3}, we derive for \pks\ $M_{\rm BH}\simeq f_G~3\times 10^8 ~M_\odot$. Although the inclination angle of \pks\ is unknown, the nearly symmetric
position of the lobe-like radio structures suggests that the system is not seen
pole-on. If we conservatively assume $i=30^o - 45^o$, then $f_G\simeq 0.4-0.5$,
leading to a mass estimate  of $M_{\rm BH}\simeq 1.5 \times 10^8 ~M_\odot$.

In summary, all the optically-independent methods consistently yield  
$M_{\rm BH}$ values of the order of few units in $10^8 ~M_\odot$, which is about
a factor $\sim$5 larger than the value derived from the virial theorem. Interestingly,
this is in agreement with the corrective factor proposed by \citet{marco08}.
However a systematic study of the $M_{\rm BH}$ distribution of well-defined samples of 
NLS1s and BLS1s obtained with optically-independent methods
is necessary before drawing any general conclusion. Considering all four methods,
$M_{\rm BH}$ in \pks\ ranges between $6 \times 10^7~-~ 3 \times 10^8 ~M_\odot$ with a 
mean of $1.8 \times 10^8 ~M_\odot$.

\section{A multi-wavelength view of \pks}
\subsection{Short-term variability}
\begin{figure}
\includegraphics[bb=70 25 475 675,clip=,angle=0,width=8.5cm]{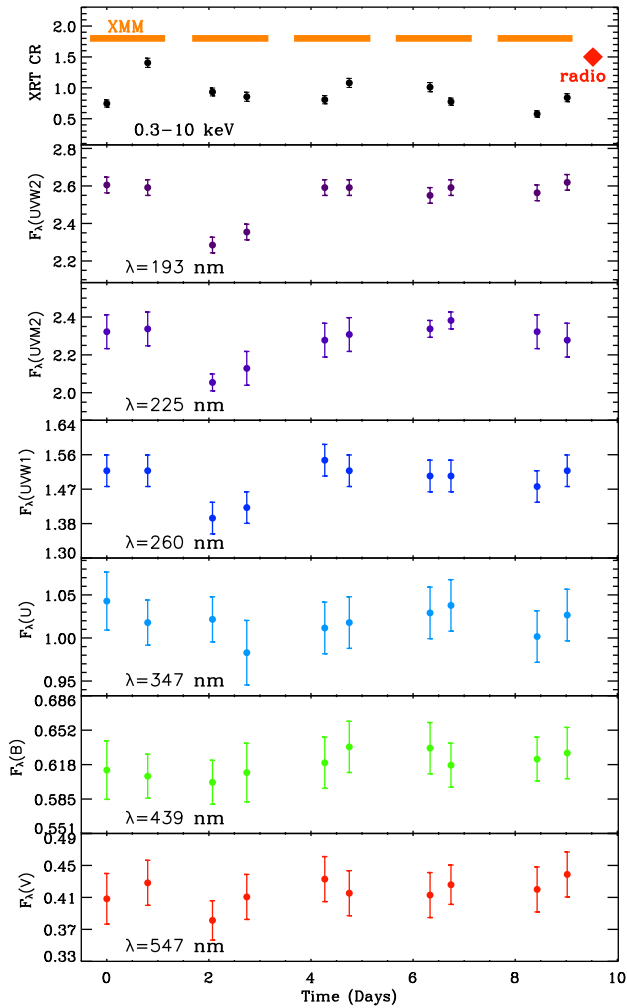}
\caption{\scriptsize
\swift\ XRT and UVOT light curves of \pks\ during the simultaneous campaign
with \xmm\ in September 2008. The UVOT fluxes are in units of 
$10^{-14}~{\rm erg~cm^{-2}s^{-1} \AA^{-1}}$ and are corrected for Galactic absorption.
The duration of the five \xmm\ orbits is shown in the top panel as well as the occurrence
of the \atca\ observation.}
\label{figure:fig4}
\end{figure}

Before investigating the properties of SEDs derived
from the multi-wavelength campaign in September 2008,
it is important to assess the presence of variability at the different 
wavelengths probed by \swift. The light curves
obtained from the six UVOT filters and from the XRT are shown
in Fig.~\ref{figure:fig4}. The plotted optical and UV time series are the
fluxes in units of $10^{-14}~{\rm erg~cm^{-2}s^{-1} \AA^{-1}}$ and are
corrected for Galactic absorption (see \citealt{rom09} for details). 
To be conservative, we have used error-bars of 3$\sigma$ to assess the presence of significant 
variability with a $\chi^2$ test (the same conclusions are obtained with 2$\sigma$ errors),
because the 1$\sigma$ statistical errors  underestimate
the true uncertainty on the optical-UV fluxes \citep[see][]{pool08}. In the 
top panel of Fig.~\ref{figure:fig4} in addition to the XRT light curve,
we show the occurrence and duration of the \xmm\ EPIC pn observations.
Sub-intervals of the duration of 2 ks simultaneous to the \swift\ pointings
were used for the SED analysis.

According to a  $\chi^2$ test no significant variability is present in the U, B, 
V bands ($P_{\chi^2}\simeq 0.9-0.95$) and only marginal variability is detected in 
the UVW1 band ($P_{\chi^2}\simeq 0.2$),
as suggested by a visual inspection of the four bottom panels of Fig.~\ref{figure:fig4}.
On the other hand, the two higher energy UV bands 
(UVM2 and UVW2) and the X-ray range all show statistically significant variability 
($P_{\chi^2} < 10^{-4}$). There are however qualitative and quantitative differences
between the UV and X-ray light curves: the former show only moderate variability,
$F_{\rm var,M2}=(3.0\pm1.4)$\%, $F_{\rm var,W2}=(4.1\pm0.5)$\% (where $F_{\rm var}$
is the fractional variability that measures the normalized variance corrected for
the statistical uncertainties), which appears to
be associated with only two data points (specifically the third and fourth points
in Fig.~\ref{figure:fig4}) that are 
significantly lower than the others. On the other hand, the XRT light curve shows strong
variability ($F_{\rm var,XRT}=(28.6\pm0.9)$\%) throughout the entire 8-day interval.

In conclusion, the \swift\ campaign in September 2008 confirms the presence 
of strong variability in the X-ray band of \pks\ and reveals the presence of 
moderate variability
in the UV bands, whereas the optical bands appear to be consistent with the
hypothesis of constant flux over short timescales.

\subsection{Broad-band SED}
One of the best ways to get insights into the energetic of the central engine
in \pks\ and in AGNs in general is to study the broadband SED, 
as demonstrated by numerous successful past 
studies \citep[e.g. see][]{elv94}. However, one of the major problems in this
field has been the lack of truly contemporaneous observations in different 
energy bands, which could lead to inaccurate results in virtue of the fast
variability of AGNs. Recently this kind of studies has markedly improved 
with the advent of 
\swift\ thanks to its ability of observing simultaneously in several energy bands
coupled with its highly flexible schedule.

Here we analyze the contemporaneous SEDs of \pks\ obtained with the \swift\ UVOT and \xmm\
EPIC pn during a simultaneous campaign
in September 2008. The choice of using 2000 s segments of the EPIC pn instead
of the \swift\ XRT data is dictated by the much higher throughput of the EPIC camera
that allows a more detailed spectral analysis of the X-ray spectrum.

Following the procedure adopted by \citet{vasu09}, we converted 
the UVOT fluxes using the \verb+FLX2XSP+ command in \verb+FTOOLS+. 
The resulting spectra were then combined with the 0.3--10 keV EPIC spectra and
fitted in \verb+XSPEC+ with a model that comprises 
a disk and two Comptonization components: {\it DISKPN+WABS(BMC+BMC)}. Only the 
Comptonization models are absorbed by the Galactic $N_{\rm H}$, because the 
optical and UV data were already corrected for absorption. 

As explained in $\S4$
the BMC is a simple but comprehensive Comptonization model that can fit both
thermal and bulk Comptonization processes. The use of a
Comptonization model to parametrize the soft excess below 2 keV is guided 
by a thorough spectral analysis of the highest quality \xmm\ spectra, which suggests
that this component may arise from a hot "skin" in the innermost part of the disk
in objects that accrete at rates close or above the Eddington limit \citep{papa10a}. 
For the hard X-rays we also used a BMC model instead of the phenomenological
power law model (PL) because the BMC parameters are computed in a self-consistent 
way, and, unlike the PL, the power 
law produced by BMC does not extend to arbitrarily low energies and thus does 
not affect the normalization of the thermal component nor the amount of local 
absorption.

The DISKPN model \citep{gierl99}
has three parameters: $T_{\rm max}$, $R_{\rm in}$, and the normalization
that depends on the black hole mass, the distance, the inclination angle $i$, and 
the color factor 
$\beta$. The latter two quantities, which are unknown for \pks, were kept fixed to  
$i=0^o$ and $\beta=1$, since they only marginally affect  
the luminosity derived from the direct integration of the SED \citep[see][]{vasu09}. 
In order to determine the maximum temperature of the accretion disk, we first
fitted the UVOT data only with the DISKPN model, fixing $R_{\rm in}=6~R_{\rm G}$ and
the normalization to two different values corresponding to 
$M_{\rm BH}=6\times10^7{~\rm M_\odot}$ and $M_{\rm BH}=2.5\times10^8{~\rm M_\odot}$,
respectively. The resulting best fit temperatures were respectively $\sim$ 23  eV and 
$\sim$ 8  eV. If the normalization of the DISKPN model is left free to vary,
the resulting value derived for the  BH mass is $M_{\rm BH}=2.5-2.7\times10^8{~\rm M_\odot}$,
which in good agreement with the values inferred from the three model-independent techniques
described in Sect. 4.

We then fitted the X-ray spectra by linking the temperature of seed photons 
of both Comptonization models to $T_{\rm max}$. The resulting best fit parameters for the 
BMC components and the corresponding luminosities are reported in Table 3 for the
conservative assumption that $M_{\rm BH}=2.5\times10^8{~\rm M_\odot}$. The values of
the reduced $\chi^2$ for the X-ray spectral fits  range between 0.9 and 1.2, whereas 
for the overall broadband fits $\chi^2_r$ are of the order of 1.4--1.5.
Note that the BMC parameters and the X-ray luminosity are almost insensitive of the choice 
of $kT$, whereas the bolometric luminosity $L_{\rm bol}\simeq L_{\rm 0.01-100 ~keV}$ is significantly affected by this choice: $L_{\rm bol}$ increases by a factor of $\sim$3
when $kT=23$ eV (i.e., for $M_{\rm BH}=6\times10^7{~\rm M_\odot}$). 

\begin{figure}
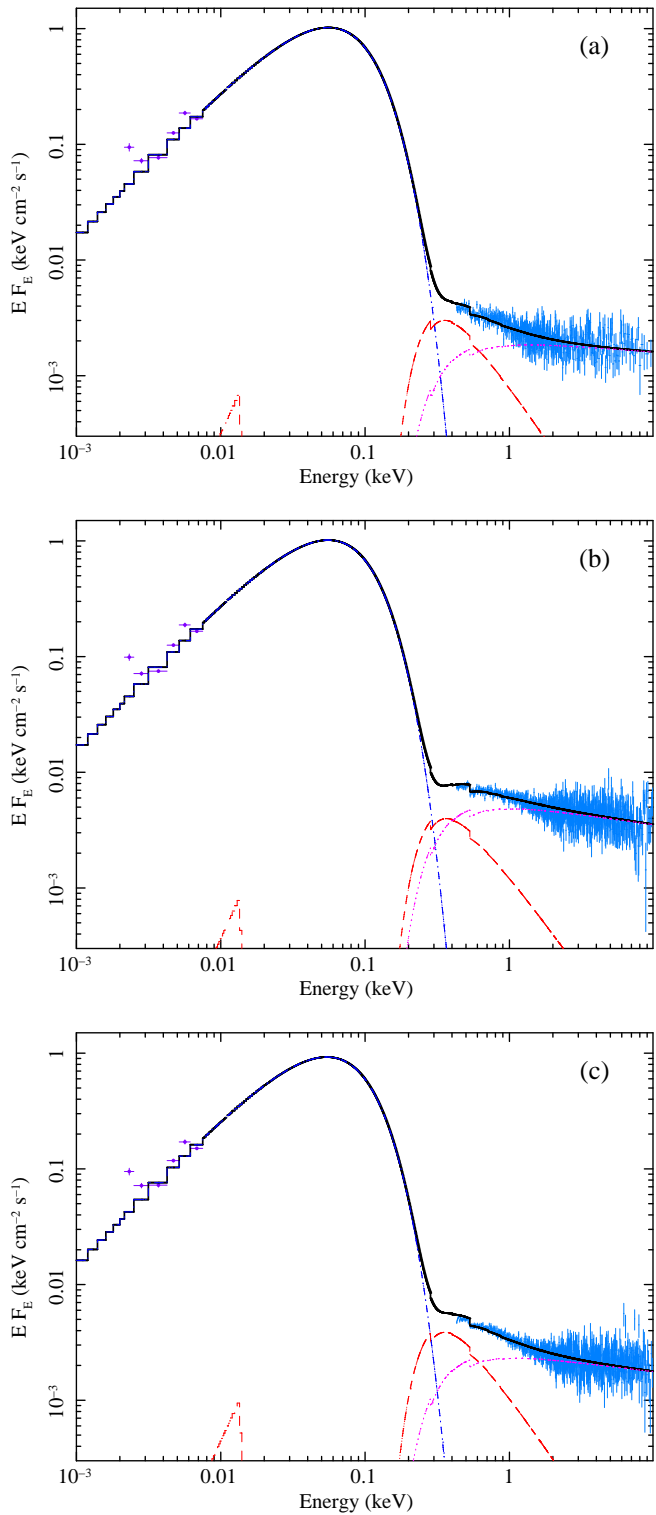

\includegraphics[bb=60 10 595 720,clip=,angle=-90,width=9.cm]{f5a.eps}

\includegraphics[bb=60 10 595 720,clip=,angle=-90,width=9.cm]{f5b.eps}

\includegraphics[bb=60 10 595 720,clip=,angle=-90,width=9.cm]{f5c.eps}
\caption{\scriptsize
Deconvolved EPIC pn spectra in the 0.4--10 keV energy range,
combined with UVOT data. The overall SEDs are fitted with a diskpn model to parametrize the optical-UV emission and two BMC models to characterize the X-ray spectrum. 
The BMC models are absorbed by Galactic $N_{\rm H}$, whereas the optical/UV data have been
already corrected for Galactic absorption. The spectrum in the top panel (a) 
represents the case of average optical-UV flux and average X-ray emission. The middle
panel (b) illustrates the case of high X-ray emission, 
whereas the bottom panel (c) represents the case of relatively low optical/UV emission.
}
\label{figure:fig5}
\end{figure}

Since a very detailed X-ray spectral analysis has been already performed and reported 
elsewhere \citep[see][]{papa10a}, here we focus on the broadband SED.
Figure~\ref{figure:fig5} illustrates three different SEDs corresponding respectively
to (a) segment 001 in the top panel, which represents the case of average optical-UV 
flux and average X-ray emission; (b) segment 002 in the middle panel, which refers    
the highest X-ray emission level; and  (c) segment 003
in the bottom panel that represents the case of low optical/UV emission.
Whereas the X-ray emission varies considerably from day to day with luminosity
changes up to a factor of $\sim$2, the changes in the optical/UV band are barely 
noticeable and correspond to variations of $L_{\rm bol}$ lower than 10\%.

In conclusion, the overall SEDs of \pks\ during the September 2008 campaign are
reasonably well fitted with a disk model parameterizing the optical/UV emission and two 
Comptonization components to describe the X-ray spectra. The SED appears to be dominated
by a nearly constant disk component, with a variable X-ray contribution which represents
about 1/100 of the $L_{\rm bol}$ with the assumption that 
$M_{\rm BH}=2.5\times10^8{~\rm M_\odot}$. The X-ray contribution is further reduced by a factor 
of $\sim$3 if $M_{\rm BH}=6\times10^7{~\rm M_\odot}$.

\begin{table*}
\scriptsize
\caption{X-ray spectral parameters and luminosity of PKS 0558-504}
\begin{center}
\begin{tabular}{lccccccc}
\hline        
\hline
\noalign{\smallskip}
Segment &   $\alpha_1$ & $\log A_1$ & $\alpha_2$ & $\log A_2$ & $L_{\rm 0.4-2~keV}$  & $L_{\rm 2-10~keV}$ &  $L_{\rm 0.001-100~keV}$ \\
  &   &  &  &  & ($10^{44}{\rm erg~s^{-1}}$) & ($10^{44}{\rm erg~s^{-1}}$) & ($10^{46}{\rm erg~s^{-1}}$) \\
\noalign{\smallskip}
\hline
\noalign{\smallskip}
001  & 2.86$^{+0.07}_{-0.10}$ & 2 & 1.09$^{+0.07}_{-0.09}$ & 0.48& 4.8  & 2.3 & 5.7 \\
\noalign{\smallskip}
002  & 2.70$^{+0.05}_{-0.08}$ & 2 & 1.17$^{+0.04}_{-0.05}$ & 0.64& 10.0 & 5.3 & 5.7 \\
\noalign{\smallskip}
003  & 2.83$^{+0.04}_{-0.04}$ & 2 & 1.15$^{+0.04}_{-0.05}$ & 0.59& 6.3 & 2.9 & 5.1 \\
\noalign{\smallskip}
004  & 2.70$^{+0.05}_{-0.07}$ & 2 & 1.19$^{+0.05}_{-0.04}$ & 0.69& 7.5 & 3.7 & 5.1 \\
\noalign{\smallskip}
005  & 2.68$^{+0.04}_{-0.04}$ & 2 & 1.10$^{+0.04}_{-0.04}$ & 0.50& 8.4 & 4.2 & 5.5 \\
\noalign{\smallskip}
006  & 2.63$^{+0.05}_{-0.07}$ & 2 & 1.24$^{+0.05}_{-0.05}$ & 0.86& 10.0 & 4.6 & 5.6 \\
\noalign{\smallskip}
007  & 2.61$^{+0.05}_{-0.07}$ & 2 & 1.17$^{+0.04}_{-0.05}$ & 0.64& 7.5 & 4.0 & 5.8 \\
\noalign{\smallskip}
008  & 2.72$^{+0.04}_{-0.05}$ & 2 & 1.10$^{+0.03}_{-0.04}$ & 0.49& 6.1 & 3.3 & 6.0 \\
\noalign{\smallskip}
009  & 2.74$^{+0.07}_{-0.07}$ & 0.7 & 1.19$^{+0.08}_{-0.06}$ & 0.70& 8.4 & 3.8 & 5.4 \\
\noalign{\smallskip}
010  & 2.81$^{+0.07}_{-0.07}$ & 2 & 1.09$^{+0.08}_{-0.06}$ & 0.69& 7.4 & 3.3 & 5.8\\
\noalign{\smallskip}
\hline        
\hline
\end{tabular}
\end{center}
\label{tab3}
\scriptsize{
{\bf Columns Table 3}: 1= Segment of the \swift\ campaign. 2= Spectral index of the BMC
model fitting the soft excess below 2 keV. 3= Measure of the Comptonization fraction
($f=a/(1+A)$) for the soft excess. 4= Spectral index of the BMC model for the 2--10 keV
range. 5= Measure of the Comptonization fraction for the 2--10 keV range. 6= Soft X-ray
luminosity. 7= Hard X-ray luminosity. 8= Bolometric luminosity\\
{\bf Notes}:
The bolometric luminosity is obtained by assuming $M_{\rm BH}=2.5\times10^8~{\rm M_\odot}$,
which leads to a best fit temperature of $kT\simeq 8$ eV. If $M_{\rm BH}=
6\times10^7~{\rm M_\odot}$, then $kT\simeq 23$ eV and the bolometric luminosity is a factor
$\sim 3$ higher than the values reported in the Table.}
\end{table*}

\section{Discussion} 

\subsection{Jet contribution}
The main goal of this work is to investigate the energetic of the radio-loud
NLS1 \pks\ and shed some light on the nature of its central engine. 
To this end, it is 
first necessary to assess the role played by the jet and more specifically whether 
the jet may dominate the high energy emission
of \pks. The findings based on the \atca\ observations reported
in $\S3$ -- the radio-loudness measurements consistent with radio-quiet Seyfert galaxies
and the extended and symmetric FRI-like radio structure which is at odds with the compact
radio emission typical of very radio-loud NLS1s-- disfavor a jet-dominated
scenario, which instead appears to be the most likely scenario for the
very radio-loud NLS1s recently detected with the {\it Fermi/LAT}.

This preliminary conclusion is supported by several independent pieces of evidence
from studies at higher energies, which can be summarized as follows: (1) the X-ray 
spectral and temporal variability appears to be fully consistent with the typical 
behavior of radio-quiet Seyfert galaxies \citep{papa10a,papa10b}; (2) the long-term 
spectral variability in the 2--15 keV
energy band is inconsistent with the characteristic trend observed in blazars with \rxte\
\citep{glioz07}; (3) unlike highly beamed AGNs, \pks\ is not detected at very high 
energy neither in the GeV range (the first year Fermi LAT catalog) nor at TeV energies
(HESS; Giebels private communication). Combining all the information available for \pks\
it is thus reasonable to conclude that the jet emission does not play a relevant role
beyond the radio band.  Therefore, for the rest of the paper we will assume that the 
bulk of the optical-UV-X-ray emission is associate with the accretion flow only.

\subsection{Energetic of \pks}
The main result from the analysis of the broadband SED  (described in $\S5$) is that 
the emission 
of \pks\ is dominated by a weakly variable UV bump ($F_{\rm var,UV}\simeq 4\%$),
which is commonly associated with the emission from the accretion disk. 
Conversely, the X-ray radiation, which we have interpreted as emission
from two Comptonization components, appears to be highly variable ($F_{\rm var,X}\simeq 30\%$)
but encompasses only a small fraction of the total emission. It is instructive to
assess in a more quantitative way which of the variable components (UV or X-rays) is energetically more important. Since $F_{\rm var}$ provides a measurement of the 
variability normalized over the average flux, a measurement of the ``true" variability
(hereafter $var_{\rm UV}$ and $var_{\rm X}$, respectively) can be
obtained by multiplying $F_{\rm var}$ by the respective average flux.
This test yields $var_{\rm X}/var_{\rm UV}=1.3$, indicating that the X-ray
component is the dominant variable component.

In order to get helpful insights into the nature of the central engine in \pks\ it
is crucial to constrain its accretion rate in term of Eddington units. This information
can be readily obtained by knowing the mass of the central black hole and measuring the
bolometric luminosity.  Depending on the method used (see $\S4$) the black hole mass of 
\pks\  may range between $6\times10^7{\rm ~M_\odot}$ and  $3\times10^8{\rm ~M_\odot}$. 
To be conservative, in the rest of the paper we will assume  
$M_{\rm BH}=2.5\times 10^8{\rm ~M_\odot}$ which corresponds to an Eddington luminosity
of $L_{\rm Edd}=3.25\times 10^{46}$ \lum\ (note that this choice of $M_{\rm BH}$
does not affect the main findings nor our conclusions). Combining this value with the 
average bolometric
luminosity derived from the direct integration of the broadband SED (see Table 3) leads to
an accretion rate of $\lambda_{\rm Edd}\equiv L_{\rm bol}/L_{\rm Edd}\simeq 1.7$ that strongly suggests that \pks\ accretes at super-Eddington rate. 
Note that with the less conservative assumption of
$M_{\rm BH}=6\times 10^7{\rm ~M_\odot}$ (which corresponds to a lower Eddington luminosity
and a higher bolometric luminosity), the accretion rate in Eddington units would be of
the order of $\lambda_{\rm Edd}\simeq 25$.

Having measured the X-ray and bolometric luminosity over a period of ten days, it is
possible to compute not only the average bolometric correction $\kappa_{\rm 2-10~keV} \equiv 
L_{\rm 2-10~keV}/L_{\rm bol}$ but also to investigate if it varies over time. From 
Table 3 we infer that $\kappa_{\rm 2-10~keV}$ indeed varies by a factor of $\sim$2.5
ranging between 108 and 249, with an
average value of $162\pm39$ (where the error quoted is 1$\sigma$). The large value
of $\kappa_{\rm 2-10~keV}$ appears to be in qualitative agreement with recent 
studies that suggest that the 
X-ray bolometric correction is directly proportional to the Eddington ratio 
\citep[e.g.][]{vasu07,vasu09}. A more quantitative comparison can 
be performed using the recent results presented by \citet{lusso10}, which are based on 
the optical and X-ray
study of a sample of 545 X-ray selected type 1 AGNs from the XMM-COSMOS survey. More specifically,
plugging $\lambda_{\rm Edd}=1.7$  (the value of the Eddington ratio derived for \pks) into
equation (14) from \citet{lusso10} we derive $\kappa_{\rm 2-10~keV}=151_{-20}^{+24}$,
which is fully consistent with the average value empirically derived from the SED. 

Since \pks\ possesses a radio jet,
it is instructive to estimate the the inflow accretion rate
and try to compare it to the the outflow mass rate associated with the jet,
and similarly compare the accretion luminosity to the kinetic power. 
With the reasonable assumption that 
$L_{\rm bol}=L_{\rm accr}=\eta\dot M_{\rm in} c^2$, we derive an accretion inflow rate of 
$\dot M_{\rm in} \simeq 1 {~\rm M_\odot/yr}$, assuming a typical radiative efficiency of $\eta=0.1$
(larger values of $\dot M$ are obtained if photon trapping effects play a significant
role causing $\eta$ to decrease considerably).
Following \citet{merlo07} it is possible to derive an estimate of the kinetic power 
of the jet for radio galaxies by making use of
the formula $\log L_{\rm kin}=0.81\log L_{\rm R} +11.9$, where $L_{\rm R}$ is the
nuclear radio luminosity at 5 GHz. Using the  value measured with the \atca,  
$L_{\rm 5 GHz}=1.7\times10^{41}$ \lum, we  derive a kinetic power of the order
of $L_{\rm kin}\simeq 2\times10^{45}$ \lum. The latter is considerably lower than the luminosity
associated with the accretion flow, which is of the order of $5\times10^{46}$ \lum.
We can estimate the outflow mass rate from $L_{\rm kin}=(\Gamma-1)\dot M_{\rm out}c^2$
by assuming an appropriate value for the bulk Lorentz factor. With $\Gamma=3$, which is
a typical value for radio galaxies and consistent with the value inferred from the brightness temperature derived from the radio variability (Kedziora-Chudczer et al. 2010, in 
preparation), we derive  
$\dot M_{\rm out} \simeq 2\times10^{-2} {~\rm M_\odot/yr}$, which is about two orders
of magnitude lower than the value inferred for the accretion rate $\dot M_{\rm in}$.

\subsection{Analogy with GBHs} 
Our recent study based on a 1-year \rxte\ monitoring campaign suggests that the spectral
variability behavior of \pks\ in the 2--15 keV energy band mimics GBHs in their highly 
accreting IS \citep{glioz07}. We can now test this analogy in a more quantitative
way. First, we can compare the results from the energetic section relative to the disk-jet
interaction with the findings of \citet{osu09} that are derived from radiation-magneto-hydrodynamic simulations of a stellar BH system accreting at the Eddington rate. 
The general agreement with the simulations
(specifically, $\dot M_{\rm out}\ll \dot M_{\rm in}$ and $L_{\rm kin} < L_{\rm accr}$)
lends further support to the hypothesis that \pks\ is a large scale analog of a GBH
in a highly accreting state.

The analogy with the IS/VHS is supported by the direct comparison of \pks\
with GRO J1655-40 in the $\Gamma - N_{\rm BMC}$ diagram that we used to estimate the 
black hole mass (see Fig.~\ref{figure:fig3}). Specifically, the common
spectral variability trend occurs when GRO J1655-40 is switching from the LS 
to the HS (i.e., the source is indeed in the IS) during the rise of the outburst in
March 2005 (see details in \citealt{shap07}). Interestingly, the X-ray outburst of 
GRO J1655-40 was accompanied by enhanced emission in the radio band that 
faded before the
peak of the X-ray outburst. Based on the values reported by \citet{shap07},
the ratio between radio/X-ray and radio/optical fluxes places GRO J1655-40 in the same lower-left quadrant of the radio-loudness plot shown in Fig.~\ref{figure:fig2}
for \pks. 

Further support to this conclusion comes from the comparison of the broadband 
spectral index $\alpha_{\rm OX}$ of \pks\ with the corresponding quantity $\alpha_{\rm GBH}$,
which was recently introduced by \citet{sobo09} to characterize the broadband
 spectrum of GBHs and allows a direct comparison with AGNs.
During our short multi-wavelength campaign in October 2008, $\alpha_{\rm OX}$
ranged between 1.25--1.4 while the photon index varied from 2.1--2.25. Importantly,
this is the very range corresponding to the intermediate state in GBHs (see Figure 6
in \citealt{sobo09}).

Finally, it is instructive to compare \pks\ to GRS 1915+105, which is thought to be the only
microquasar that regularly accretes at or above the Eddington limit. Interestingly,
during its spectral transition hard-to-soft the X-ray spectra of GRS 1915+105 are
well fitted by two BMC components (a hard component with $\Gamma_1=$1.7--3 and a soft
component with $\Gamma_1=$2.7--4.2) which evolve following the trend shown in
Fig.~\ref{figure:fig3} \citep{tita09}. Also for GRS 1915+105 the location
of the IS/VHS in the $\Gamma - N_{\rm BMC}$ diagram is consistent with the position of
\pks\ (see Figure 13 of \citealt{tita09}). 

\subsection{Conclusion}
In conclusion, our multiwavelength analysis of the emission properties of \pks\
ranging between the radio and the hard X-rays reveals that this source has a radio
jet, yet it  is accreting at the Eddington level or above.
All our findings appear to independently lend support to the hypothesis that \pks\  
may be a large scale analog of the IS observed in GBHs. But can we extend this 
analogy to the whole class of NLS1s? In other words, can \pks\ be considered
a prototype of NLS1s? At first sight, the ``radio-loudness" of \pks\ seems to set 
this source in a small sub-class of radio-loud NLS1. However, 
 $R=10$ does not represent a strict boundary for the radio-loudness; indeed
the radio-loudness of \pks\ appears to be fully consistent with that of radio-quiet
Seyfert galaxies according to both $R$ and $R_x$ (see Fig.~\ref{figure:fig2}).
Bearing in mind that \pks\ is by far the brightest NLS1 in the X-ray band and that 
its X-ray luminosity is nearly
100 times the typical value of ``normal" radio-quiet NLS1s, similar values of the
order of $R_x \la 10^{-3}$
for ``normal" NLS1s would correspond to radio fluxes of the order of 1 mJy or less.
This 
would explain why only a small minority of NLS1s is detected in the radio band. 
Additionally, it must be pointed out that in the IS the radio ejections are transients,
therefore, genuine radio-quiet NLS1 might still represent large scale analogs of the IS.

It is also interesting to compare \pks\ with the radio-loud NLS1s recently detected 
at $\gamma$-rays. The latter are characterized by large values of the radio-loudness
($R > 50$), by compact radio emission, and by SEDs that are dominated by two broad bumps
typical of jet-dominated sources. Indeed, it has been argued that these very radio-loud
NLS1s galaxies represent a the third class of $\gamma$-ray emitting AGNs along with
blazars and radio galaxies. In this perspective, \pks\ can play a crucial role since
it may represent the prototype of NLS1s with a jet seen at large viewing angles. In other
words, \pks\ may be the the first representative of non-jet-dominated radio-loud
NLS1s which are
the parent population of the very radio-loud NLS1s detected by the {\it Fermi/LAT},
similarly to radio galaxies that are considered the non-beamed parent population
of blazars.

In summary, we can hypothesize that \pks\ is indeed a prototypical NLS1.
Since the jet appears to dominate  only in the radio band, the emission 
associated with the accretion flow can be cleanly separated by that related to the jet.
This makes \pks\ an ideal system to study the disk-jet interaction. In this framework,
important insights into the physical conditions leading to the formation of jets
in highly accreting BH systems can be obtained by a systematic comparison of the spectral
and temporal properties of \pks\ with the corresponding properties of other highly
accreting NLS1 that do not show any evidence of radio jets
as well as with NLS1s that are jet-dominated. We plan to pursue this
approach in our future work.

\begin{acknowledgments} 
We that the anonymous referee for the constructive suggestions that have improved
the clarity of the paper.
MG acknowledges support by the NASA SWIFT (NNX08AU04G) and XMM-Newton (NNX08AX82G)
Guest Investigator Programs and by the NASA ADP grant NNXlOAD51G.  DG
was supported by the NASA ADP grant (NNX07AH67G) and the NASA PSU contract
(Swift is supported at Penn State by NASA contract NAS5-00136).
IEP and WPB acknowledge partial support from the EU ToK grant
39965 and FP7-REGPOT 206469.
\end{acknowledgments}

\end{document}